\documentclass{article}

\begin{document}

\def\giorno{19/12/2001}

\def\sec#1{\bigskip\bigskip \noindent{\bf #1} \bigskip}
\def\subsec#1{\medskip ~~~~ {\bf #1} \medskip \par}

\def\R{{\bf R}}
\def\C{{\bf C}}
\def\H{{\cal H}}
\def\G{{\cal G}}
\def\T{{\rm T}}

\def\L{{\cal L}}
\def\M{{\cal M}}

\def\^#1{{\widehat #1}}
\def\~#1{{\widetilde #1}}

\def\sse{\subseteq}
\def\ss{\subset}

\def\pa{\partial}
\def\a{\alpha}
\def\b{\beta}
\def\eps{\varepsilon}
\def\la{\lambda}
\def\phi{\varphi}
\def\om{\omega}

{}

\title{{\bf Poincar\'e normal forms and simple compact Lie groups}}

\author{Giuseppe Gaeta\footnote{Work supported in part by ``Fondazione CARIPLO per la ricerca scientifica'' under project ``Teoria delle perturbazioni per sistemi con simmetria''.} \\ 
{\it Dipartimento di Matematica, Universit\'a di Milano} \\ 
{\it v. Saldini 50, I--20133 Milano (Italy)} \\ 
{\tt giuseppe.gaeta@roma1.infn.it} }
\date{\giorno}

\maketitle

\noindent
{\bf Summary.} {We classify the possible behaviour of
Poincar\'e-Dulac normal forms for dynamical systems in $R^n$ with
nonvanishing linear part and which are equivariant under (the fundamental representation of) all the simple compact Lie algebras and thus the corresponding simple compact Lie groups. The ``renormalized forms'' (in the sense of \cite{Ga2}) of these systems is also discussed; in this way we are able to simplify the classification and moreover to analyze systems with zero linear part. We also briefly discuss the convergence of the normalizing transformations.}

\bigskip

\noindent{\tt P.A.C.S. numbers: 02.20.-a , 02.30.Hq , 05.45.-a }

\noindent{\tt Mathematical Subjects Classification: 34C20 ,
58F36 , 70K45 } 

\noindent{\tt Keywords: Normal forms, Symmetry, Compact simple Lie groups} 

\noindent{\tt Last modified: \giorno}

\bigskip\bigskip\bigskip

\section{Introduction.}

The theory of {\it normal forms}, created by Poincar\'e one century ago, is still one of the most powerful tools at our disposal to investigate nonlinear systems, both theoretically and in dealing with concrete applications.

In dealing with Poincar\'e-Dulac (and, in the hamiltonian case, with Birkhoff-Gus\-tav\-son) normal forms \cite{Arn,AIl,Be1,Bru,Gle,IAd,Ver,Wal}, the presence of symmetries introduces some peculiar features and even make generic some  situations which in the non-symmetric case are highly exceptional. 

Normal form in the presence of linear symmetries were studied by
Belitskii \cite{Be2} and, indipendently, by Broer \cite{Bro}, Takens \cite{BrT}, and by Elphick {\it et al.} \cite{Elp} who completely clarified the issue in general terms; see also the very clear exposition given by Iooss and Adelmeyer in \cite{IAd}. The theory was subsequently extended to consider nonlinear group actions as well \cite{CG2,CG3}. 
 
In relatively recent years it has been realized that the presence of
symmetries has also deep consequences on the convergence properties of the (in general, only formal) normalizing transformation; in particular, they can ensure its convergence \cite{BrW,Cic,CWt,Mar,Sto,Wal2}.

For a general discussion of normal forms theory in the presence of
symmetries, we refer to \cite{CG3}; a shorter treatment is given in \cite{CG2}.
 
\medskip

However -- maybe surprisingly in view of this stage of developement -- the theory of normal forms in the presence of symmetries has never been applied to investigate systematically general vector fields and dynamical systems with symmetries of physical interest.

In this note I want to focus on the case, whose relevance for Physics does not need to be emphasized, where symmetries are described by a {\it compact simple Lie group}\footnote{The mathematically oriented reader will notice that our considerations would to some extent apply to a more general class of groups and algebras; we prefer not to discuss this point here, but see however remark 7.} or more precisely by its Lie algebra.
  
I will assume that the reader has some basic knowledge of normal forms theory, e.g. as given in \cite{Gle,IAd,Ver}; I will also assume knowledge of basic group theory, see e.g. \cite{Kir,NaS}.

\medskip

I will apply the general theory to obtain a classification of normal forms for systems equivariant under the fundamental (defining) representations of simple Lie algebras and thus in particular under the fundamental orthogonal representation of simple, including classical, Lie groups.

I will also briefly discuss the convergence of the transformation to normal form (normalizing transformation) for these systems, solely on the basis of the symmetry properties and of the linear part (assumed to be nonzero) of the system. It should however be recalled that when dealing with normal forms,
formal results are also very useful in practice: in concrete cases the
normalization is performed only up to some finite order $N$, then we can study the truncation of the system at this order and consider the higher order terms as perturbation, to control by some other perturbative techniques; see e.g. \cite{CG3,Gle,IAd,Ver}.

The standard normal forms theory is based on the action of an operator (the homological operator) associated to the linear part of the system, and is obviously not able to deal with systems
having zero linear part; thus in sections 3 and 4 we will assume that the linear part of the system is nonvanishing. However, ``renormalized forms'' are also able to deal with systems having zero linear part, and they will be considered, with this approach, in section 5. As far as I know, this is the first application of normal forms theory to systems which do not have a linear part at singular points allowing such general results. (Generally speaking, normal forms are of scarce effectiveness at non-regular singular points; here the symmetry comes to help us).

\bigskip

The setting to be considered is that of dynamical systems (vector
fields, formal power series) in $\R^n$ having a fixed point (a zero) in the origin; this of course is also representative of a more general
situation, where we work in a local chart in the neighbourhood of a
fixed point for a dynamical system (vector field) defined on a smooth
$n$-dimensional manifold $M^n$. This dynamical system is written
locally in terms of a (formal) power series expansion as 
$$ {\dot x} \ = \ f (x) \ = \ \sum_{k=0}^\infty \, f_k (x) \ , \eqno(1) $$ 
where $x \in \R^n$, $f : \R^n \to \T \R^n \approx \R^n$, and $f_k (x)$ is homogeneous of degree $k+1$ in $x$, i.e. $ f_k (ax) \ = \ a^{(k+1)} \, f_k (x)$. 
We can equivalently consider the vector field (VF) $X \equiv X_f$ given in the $x$ coordinates by  $$ X_f \ := \ f^i (x) \, { \pa ~\over \pa x^i} \ . \eqno(2) $$

Notice that $f_k$ is homogeneous of degree $k+1$ rather than
$k$; this notation is justified on the one hand by the action of $f_k$ under the fundamental Lie-Poisson bracket\footnote{The homological operator ${\cal L}$ is defined in terms of this as ${\cal L} (.) := \{ f_0 , . \}$.} $\{.,.\}$ defined by  $ \{\phi,\psi\}^i := (\phi^k \cdot \pa_k ) \psi^i - (\psi^k \cdot \pa_k ) \phi^i$, so that $\{ f_k , . \} : V_m \to V_{m+k}$, where $V_m$ is the vector space of vector polynomial homogeneous of order $m+1$ in the $x$; and on the other hand by the fact that if we rescale coordinates by $x = \eps \xi$ (so that $\xi$ is of
order one in the neighbourhood of the origin of order $\eps$ we will have
to consider), then (1) will read as 
$ {\dot \xi} \ = \ \sum_{k=0}^\infty \, \eps^k
\, f_k (\xi)$. 
\medskip

{\it Remark 1.} The bracket $ \{\phi,\psi\} $ introduced above is nothing else than a translation of the familiar commutator of vector fields in terms of components. That is, if $[X_f , X_g ] = X_h$, we have $h = \{ f, g \}$. $\odot$

\section{Systems equivariant under simple Lie algebras and
classical Lie groups.}

We will then consider a simple Lie algebra $\G$ and the (simply
connected, compact and actually orthogonal\footnote{We recall that for
each simple Lie algebra $\G$ there exists such a group admitting $\G$ as the group algebra (Ado's theorem) \cite{Kir,NaS}.}) simple Lie group $G$ obtained from this by exponentiation; we consider for these the fundamental (defining) matrix representation in $\R^n$; this  representation of $G$ in $\R^n$ will be
denoted by $T = \{T_g,g \in G \}$, where $T_g$ are matrices. 
The Lie algebra $\G$ will also act via matrices; we will denote a basis of $\G$ by $\{ H_{(1)} ,..., H_{(r)}
 \}$. Equivalently, $\G$ is spanned by vector fields $\{ Y_1 ,
... , Y_r \}$ given in $x$ coordinates by $Y_\a = \phi^i_\a (x) (\pa / \pa x^i)$ with $\phi^i_\a (x) = (H_{(\a)})^i_k x^k $.

A (possibly formal) dynamical system of the form (1) [vector field of
the form (2), power series of the form in the r.h.s. of (1)] is
$G$-equivariant if $\forall g \in G$ and $\forall x \in \R^n$, $ f (T_g
x)  = T_g f(x)$ (we use here linearity of the action; and also,  implicitely, the standard euclidean connection on $\R^n$ to identify tangent spaces at $x$ and at $T_g x$); this implies in particular that it is $\G$-symmetric, i.e. that  $$\{f(x), \phi_\a (x) \} \ = \ 0 \ \ \ \ \forall \a = 1,...,r \ . \eqno(3)$$ 
\medskip

{\it Remark 2.} Notice that for groups obtained by direct exponentiation of $\G$ and thus simply connected -- such as e.g. $SO(n)$ or $SU(n)$ -- the  $G$-equivariance is implied by $\G$-symmetry, while for a multiply connected group -- such as e.g. $O(n)$ or $U(n)$ -- we should still check that $f$ satisfies the equivariance relation with respect to one element of each connected component of $G$ (i.e. for $G/G^0$, with $G^0$ the connected component of the identity in $G$). Thus, for all the
$\G$-symmetric NF given below, one should still add the restriction to terms allowed by the equivariance condition under such a discrete group. $\odot$ 
\medskip

{\it Remark 3.} A situation which is also of special physical interest is that where the system is hamiltonian; or also it enjoys, besides the symmetry corresponding to the connected Lie group generated by $\G$, a {\it reversing symmetry} \cite{BHS,LaR,Sev}, or even more generally a $k$-symmetry \cite{Lam}. It is well known that in these cases the NF will also be, respectively, hamiltonian \cite{Arn,AIl,IAd} or reversible \cite{BHS,Ga1,LaR,Sev}, or $k$-symmetric \cite{Lam}. 
In the hamiltonian case, as remarked by Birkhoff, one can also 
work on the level of the Hamiltonian (i.e. a single function) rather than of the hamiltonian vector field. $\odot$ 
 
\medskip

{\it Remark 4.} As mentioned above, we will work in $\R^n$, i.e. assume these groups and algebras are acting through their fundamental {\it real} representation (thus, e.g., for $SU(m)$ groups this will be acting in $\R^{2m}$). 
It would not be to us to choose what kind of representation, over
the real or over the complex numbers should be considered: this depends on the problems we are considering, and first of all on the (physical) nature of the variables involved. However, any representation over ${\bf C}$ can be written as a representation over ${\bf R}$, so that we can in full generality just deal with the latter. $\odot$

\medskip

As stated above, we will consider simple compact Lie
groups; as well known \cite{Kir,NaS}, these reduce to the classical groups, corresponding to the simple Lie algebras $A_n$, $B_n$, $C_n$, $D_n$ (where $n \in {\bf N}$), and to the exceptional groups
corresponding to the algebras $E_6,E_7,E_8,F_4,G_2$.

The fundamental real representations of simple compact Lie groups share a common property: 
 
\medskip

\noindent 
{\bf Lemma 1.} {\it Let $G$ be a simple compact Lie group and $\G$ its Lie algebra. Then all polynomial dynamical systems which are $\G$-equivariant (equivalently, $G$-equivariant) are necessarily written in the form 
$$ {\dot x} \ = \ \sum_{p=0}^s \, \a_p (r^2 ) \, K_p x \ , \eqno(4) $$ 
where $r^2 = \sum_{i=1}^n (x^i)^2$, the $\a_p$ are power series in $r^2$, and the $K_p$ are a basis of matrices commuting with $T$.} 
 
\medskip 

\noindent 
{\it Proof.} This result can be obtained in several ways exploiting the properties of these representations, see \cite{GRo}. The simplest way is perhaps to use the fact that all these representations are orthogonal (isometric) and are transitive on the unit sphere $S^{n-1} \ss \R^n$ of their carrier space. 
Equivalently, all the $H_{(i)}$ are antisymmetric matrices, and the vectors $\eta^i = H_{(i)} x$ span the tangent space to the sphere of radius $|x|$ in $x$; see \cite{GRo} for more detail. 
Finally, one can also proceed by direct computation of the Molien functions for each representation. \hfill $\triangle$
 
\medskip

Coming back to (4), this can be expressed by saying that the $K_p$ span the
centralizer ${\cal C} (T)$ of $T$ in $GL(n,\R)$, and that $G$-equivariant
vector fields are a module over the algebra of functions of $r^2$, spanned by
matrices in ${\cal C} (T)$. 
 We will take $K_0 = I$, and $s+1 \equiv c(T)$
is the dimension of ${\cal C} (T)$.

It happens that ${\cal C} (T)$ can assume only a very limited set of
forms, and correspondingly $c(T) \in {\bf N}$ a very limited set of values; these are 
completely classified by the real version of Schur Lemma, which we
quote here in full generality from Kirillov \cite{Kir}:
\medskip
 
\noindent 
{\bf Proposition 1 (Schur Lemma).}
{\it Suppose that the dimension of the irreducible
 representation $T$ in a
linear space over the field ${\bf K}$ is no larger
 than countable. Then, if
${\bf K} = \C$, ${\cal C} (T) \simeq \C$; if
 ${\bf K} = \R$, then ${\cal C}
(T)$ is isomorphic to either $\R$ or $\C$ or
 ${\bf H}$.}
\medskip

\noindent 
{\bf Corollary.} {\it The intertwining number $c(T) = {\rm dim}_{{\bf K}} [ {\cal C}
(T) ]$ is equal to 1 for ${\bf K} = \C$ and can take the values 1 or 2 or 4
if ${\bf K} = \R$.}
\medskip

A representation over $\R$ is called of real, complex or quaternionic type according to the form of ${\cal C} (T)$. We have \cite{Kir}:
\medskip

\noindent
{\bf Proposition.} {\it Let $T_\C$ be the complexification of the real
irreducible representation $T$. If $T$ is of real or complex or quaternionic
type, then $T_\C$ is, respectively, irreducible or the sum of two
inequivalent irreducible representations or the sum of two equivalent
irreducible representations.}
\medskip

{\it Remark 5.} As well known, and as readily seen from the above
formulation, for complex representations (${\bf K} = {\bf C}$) the Schur lemma tells that ${\cal C} (T)$ reduces to (complex) multiples of the identity and thus is trivial. $\odot$
\medskip

{\it Remark 6.} The (polynomial) VFs of the form (4) are sometimes called {\it quasilinear}, corresponding in algebraic terms to a module over the set ${\cal I}_\G$ of (polynomial) invariant functions for $\G$ \cite{CG2}, which in this case (due to transitivity over 
the unit sphere) is simply given by functions of $r^2 = (x_1^2 + ... + x_n^2)$: we denote the set of quasilinear (polynomial) equivariant VFs as ${\cal Q}_\G := {\cal I}_\G \otimes {\cal M}_\G$, where ${\cal M}_\G$ are the linear equivariant VFs, ${\cal M}_\G = \{ (Mx)^i \pa_i \, : \, [M,H_i ] = 0 \ \forall i = 1,...,r \}$. Notice that dealing with normal forms we will have to consider polynomial vector fields, hence the dependence on $r^2$ instead than $r$ in (4). $\odot$
\medskip

{\it Remark 7.} Lemma 1 above can then be stated by saying that in the case of the defining representation of simple Lie algebras (and groups), and more generally of any linear representation which is transitive on the unit sphere of its
 carrier space (or equivalently satisfies the other conditions
discussed in \cite{GRo} and mentioned above), all the equivariant vector fields are quasilinear, and all the polynomial equivariant VFs are polynomial equivariant quasilinear VFs. We will actually use only the fact that (4) represents the most general equivariant VF in these cases (it is then clear that our discussion would extend to more general Lie groups and algebras). $\odot$
\medskip

{\it Remark 8.} It should be mentioned that (4) ensures the dynamical system (the VF) can be written in terms of generalized gradients \cite{GRo} in terms of $(s+1)$ scalar functions $\{ \H_0 (r^2) , ... , \H_s (r^2) \}$ as $ {\dot x}^i \ = \ \sum_{p=0}^s \,
(K_{(p)})^{ij} \, \nabla_j \, \H_p$. 
As $s+1$  is never higher than $n$, it can be convenient to deal with these scalar functions rather than with the corresponding dynamical system or VF, which are given by $n$-dimensional vectors, i.e. by $n$ scalar functions. $\odot$

\section{The Schur classification and normal forms.}

We do now focus on groups rather than algebras (for ease of language) and use the classification provided by Schur Lemma, considering separately the three possible cases. For ease of notation, we will write $\b_p := \a_p (0)$, so that the linear part of (4) reads $ f_0 (x) \ = \ A x \ = \ \sum_{p=0}^s \, \b_p \, K_p \, x$.

We will assume $A \not= 0$, as for $A =0$ the Poincar\'e 
procedure can not give any result (the case $A=0$ is discussed in section 5). 

\subsection{Case A: ${\cal C} (T) \, \simeq \, \R$.}

In this case ${\cal C} (T)$ corresponds to multiples of the identity, i.e. $s=0$, and hence (4) reduces to
$$ {\dot x} \ = \ \a_0 (r^2) \, x \ ; \eqno(5) $$
if $\b_0 = 0$ (see above), the linear part of (5) vanishes, and we cannot do anything about normalizing it; for $\b_0 \not= 0$, all the eigenvalues $\{ \la_1 ,... , \la_n \}$ are equal to $\b_0 \in \R$, and thus no resonance is present, and moreover the spectrum of $A$ belongs to a Poincar\'e domain (see section 4): thus we know \cite{Arn,AIl,Bru} that, as also recalled below, the system can be brought to its Poincar\'e normal form, which coincides with its linear part ${\dot x} = Ax$, by
a transformation which is convergent in a ball of radius $\eps$ around the origin, with $\eps > 0$.

Notice that in this case we have a gradient system, and we can equivalently Birkhoff normalize the function (potential) $\H_0 (\rho)$, where $\rho = r^2$, reducing it (in normalized coordinates $\^x$) to its
linear part in $\^\rho$; in terms of potentials, this means reducing
$\H_0$ to its quadratic part (in normalized coordinates).

The description of nonzero solutions to (5) is immediate: $x(t) = \rho (t) x(0)$, where $\rho (t)$ is a scalar function, solution to ${\dot \rho} = 2 \rho \a_0 (\rho )$ with $\rho (0) = |x(0)|^2$

{\bf Example.} An example of case A is provided by $\G = so(3) \equiv B_1$.

\subsection{Case B: ${\cal C} (T) \, \simeq \, {\bf C}$.}

Now, according to Schur Lemma, $T$ is irreeducible over ${\bf R}$ and is given by $T = T_0 \oplus \~T_0$ as a complex
representation (notice this implies $n = 2m)$; using coordinates
corresponding to this decomposition of $T$, (4) reads
$$ {\dot x} \ = \ \a_0 (r^2) \, x \ + \ \a_1 (r^2) \, J x \ , \eqno(6) $$
where $J$ is the standard simplectic matrix in dimension $n = 2 m$, representing the imaginary unit $i$, and is written
(with $I$ the $m$-dimensional identity matrix) as
$$ J \ = \ \pmatrix{ 0&- I\cr I&~0\cr } $$

The linear part is given by $ A \ = \ \b_0 I \, + \, \b_1 J$;
if $\b_0 = \b_1 = 0$, it vanishes and we cannot do anything to normalize (6); for $A\not=0$, we should consider three subcases.
\medskip

\underbar{{\it Case B1.}} If $\b_0 \not= 0$ and $\b_1 = 0$, we are in the same situation as in case A considered above: i.e. all the eigenvalues of
$A$ are real and equal to $\b_0$, so they belong to a Poincar\'e domain and
no resonance is present; the system can be normalized by a convergent
transformation (in a ball of radius $\eps > 0$), and the normal form is
linear, ${\dot w} = A w$ in normalized coordinates.
\medskip

\underbar{{\it Case B2.}} If $\b_0 \not= 0$ and $\b_1 \not= 0$ as well, the
$n=2m$ eigenvalues of $A$ split in two groups: $m$ eigenvalues are equal to
$\la_j = \b_0 + i \b_1$, and $m$ are equal to $\la_j = \b_0 - i \b_1$. In
this case the eigenvalues do still belong to a Poincar\'e domain, so that
the normalizing transformation is guaranteed to be convergent in a ball of
radius $\eps > 0$, and they cannot give origin to
resonances: in fact, any linear combination $\sum_i m_i \la_i$ of the $\la_i$ with integer coefficients $m_i$ satisfying $\sum_i m_i = m >1$, has real part equal to $m \b_0 \not= {\rm Re}(\la_j)$. Thus the NF is again linear, ${\dot w} = Aw$ in normalized coordinates.
\medskip

\underbar{{\it Case B3.}} If $\b_0 = 0$ but $\b_1 \not= 0$, the eigenvalues
of $A$ do again split in two groups, with $m$ eigenvalues $\la_j = i \b_1
\equiv i \om$, and $m$ eigenvalues $\la_j = - i \om$. Now the eigenvalues
do {\it not} belong to a Poincar\'e domain, and moreover they give origin
to an infinite number of resonances.

It is easy to check that in this case (6) is already in NF, and conversely the most general NF with linear part $A = \b_1 J$ ($\b_1 \not= 0$) is given exactly by (6). Thus, no (standard) Poincar\'e-Dulac normalization will be performed here.

Notice that here the linear part ${\dot x} = Ax$ represents a hamiltonian system, while the full system (6) is in general -- i.e. unless $\a_0 (r^2) \equiv 0$ -- non hamiltonian; the NF is also in general non-hamiltonian.

In this case we can apply to the system the ``further normalization''
procedure \cite{Ga2,GaB}; the formal computations are analogous to those for $m=1$, which in this context can be called the $SO(2)$ case, and lead to the same result\footnote{It should be stressed that \cite{Ga2} contained an incorrect result (for a degeneration of codimension at least three) for this case; the correct computation is given in \cite{GaB} and is also sketched in section 5 in a more general setting. See \cite{GLie} for a more general discussion.}: 
if $\a_0 (r^2)$ is not identically zero, we write
$\a_0 (r^2) = \sum_k a_{k} r^{2k}$, $\a_1 (r^2) = \sum_k b_k r^{2k}$ and denote by $\mu$ the smallest nonzero $k$ for which $a_k \not= 0$ (as for $b_k$, notice $b_0 \not= 0 $ in this case). Then \cite{Ga2}, 
\medskip

{\bf Lemma 2.} {\it The system can be reduced by a recursive sequence of (in general, only formal) changes of coordinates to the form
$$ {\dot x} \ = \ \b_1 J x \ + \ \left[ (c_1 r^{2\mu} + c_2 r^{4\mu} ) I \, + \, \sum_{k=0}^\mu \, d_k r^{2 k} J \right] \, x $$
with $c_i,d_k \in \R$ and $c_1 = a_\mu$.}
\medskip

It should be stressed that the convergence of the renormalizing
transformation has not been studied yet in general (see however \cite{Ga4} for partial results).  
\medskip

{\it Remark 9.} In the hamiltonian case $\a_0 (r^2) = 0$, i.e. all the $a_k$ are zero, and the renormalization procedure as described in  \cite{Ga2} is ineffective, leaving the standard normal form unchanged \cite{Ga4}; however one can still perform the ``renormalization'' by a slightly different scheme \cite{GaB}. It should also be recalled that for the hamiltonian case the same kind of result is known to hold from different considerations, and is actually a classical result \cite{SM} for $n=2$ (i.e. for one degree of freedom in hamiltonian language), recently generalized to higher dimension \cite{FM}. $\odot$
\medskip

{\it Remark 10.} Notice that in all the B cases, ${\cal C} (T)$ is abelian; this is connected to the fact that no Poincar\'e-Dulac normalization is
possible in the subcase B3, nor any ``further normalization'' is possible for
hamiltonian systems falling in this subcase. $\odot$
\medskip

{\it Remark 11.} The interpretation in terms of ``potentials'' (or
``Hamiltonians'') $\H_i$ is immediate: the system (6) can be written in terms of two of them, ${\dot x} = \nabla \H_0 (r^2) + J \nabla \H_1 (r^2)$; in
case B1 we can always normalize so that $\H_1 \equiv 0$ in normalized
coordinates, in case B2 we can always obtain $\H_1 = c \H_0$ in normalized
coordinates, and finally in case B3 we get no simplification. By further
normalization, in the last case B3 we can 
(provided $\a_0 (r^2) \not\equiv 0$) reduce $\H_0$ to the form $\H_0 = (c_1 / (2 \mu + 2)) (r^2 )^{\mu+1} + (c_2 / (4 \mu + 2))
(r^2 )^{2\mu + 1}$, and $\H_1$ to be a polynomial of order not greater than $2(\mu + 1)$. $\odot$
\medskip

{\bf Example.} An example of case B is provided by $\G = so(2) = u(1) \equiv D_1$.

\subsection{Case C: ${\cal C} (T) \, \simeq \, {\bf H}$.}

This case is actually quite rare; in particular, the only simple group 
for which the fundamental real representation is of this type, i.e.
quaternionic, is $G = SU(2)$ (with $\G = su(2) \equiv A_1$): that is, the quaternion group itself. This special case is not only needed to complete our classification, but also quite relevant mathematically and physically.

Examples of higher representations of other classical groups which are also of quaternionic type exist \cite{CG1}, but are of rather high dimension and seem to have no special (including physical) relevance, beside being not relevant to our present classification. I will therefore directly discuss the $SU(2)$ case. 

The basis matrices $\{H_1,H_2,H_3\}$ for $\G = su(2)$ will be taken to be
$$ 
H_1 = \pmatrix{0&0&1&0\cr 0&0&0&1\cr -1&0&0&0\cr 0&-1&0&0\cr } ~,~
H_2 = \pmatrix{0&0&0&-1\cr 0&0&1&0\cr 0&-1&0&0\cr 1&0&0&0\cr } ~,~ $$
$$
H_3 = \pmatrix{0&-1&0&0\cr 1&0&0&0\cr 0&0&0&1\cr 0&0&-1&0\cr } ~. 
$$

We have now four basic matrices $K_p$, with $K_0 = I$, spanning ${\cal C} (T)$. The three matrices $K_\a$ (greek indices run from 1 to 3, latin ones from 0 to 3), satisfy $K_\a^2 = - I$ and the quaternionic relations 
$ K_\a \, K_\b \ = \ \epsilon_{\a \b \gamma} K_\gamma - \delta_{\a \b} I$, and therefore
the $su(2)$ commutation and anticommutation relations $ [ K_\a , K_\b ] \ =
\ 2 \epsilon_{\a \b \gamma} K_\gamma$ and $\{ K_i , K_j \} = - 2
\delta_{ij}$. Thus, they span another
$su(2)$ algebra, not equivalent to the one spanned by the $H_\a$.

The matrices $K_\a$ can be taken to be  
$$ 
K_1 = \pmatrix{0&1&0&0\cr -1&0&0&0\cr 0&0&0&1\cr 0&0&-1&0\cr } ~,~
K_2 = \pmatrix{0&0&0&1\cr 0&0&1&0\cr 0&-1&0&0\cr -1&0&0&0\cr } ~,~ $$
$$
K_3 = \pmatrix{0&0&1&0\cr 0&0&0&-1\cr -1&0&0&0\cr 0&1&0&0\cr } ~. 
$$

Thus, the most general $SU(2)$-equivariant dynamical system (VF) expressed as a power
series is 
$$ {\dot x} \ = \ \sum_{p=0}^3 \, \a_p (r^2 ) \, K_p \, x \ ; \eqno(7) $$
the linear part of this is given by $ A \ = \ \sum_{p=0}^3 \, \b_p K_p$,
and we will write $$ \om \ = \ \sqrt{ \b_1^2 + \b_2^2 + \b_3^2} \ .$$ It is easily checked that the eigenvalues of $A$
are equal to $\b_0 \pm i \om$.
\medskip

{\it Remark 12.} By a rotation in $\R^4$ (which can actually be realized 
using the matrices $K_\a$ themselves) changing the
coordinates $x$ into coordinates $\~x$, this linear part can be reduced to a
combination
 of $K_0$ and one of the $K_\a$, say  $ A \ = \ \b_0 K_0 \, + \,
\b_1 K_1$,
 where $\~\b_0 = \b_0 $ and $\~\b_1 = \om$; the general expression
of ${\dot x} = f(x)$ in the new coordinates will be again in the form of (7). $\odot$
\medskip

As usual, if all the $\b_p$ are equal to zero, $A=0$ and there is no
Poincar\'e normalization to be considered; we will thus assume $A \not= 0$
and consider again three different subcases depending on the actual form of $A$.
\medskip

\underbar{{\it Case C1.}} If $\b_0 \not= 0$ and $\om = 0$, we are in the same situation as in cases A and B1 considered above: i.e. all the eigenvalues of $A$ are real and equal to $\b_0$, so they belong to a Poincar\'e domain and no resonances are present; the system can be normalized by a convergent transformation (in a ball of radius $\eps > 0$), and the normal form is linear, ${\dot w} = A w$ in normalized coordinates.
\medskip

\underbar{{\it Case C2.}} If $\b_0 \not= 0$ and $\om \not= 0$ as well, the $2m=4$ eigenvalues of $A$ split in two groups: $m$ eigenvalues are equal to $\la_j = \b_0 + i \om$, and $m$ are equal to $\la_j = \b_0 - i \om$. In this case the eigenvalues do still belong to a Poincar\'e domain, so that the normalizing transformation is guaranteed to be convergent in a ball of radius $\eps > 0$, and they cannot give origin to resonances, and thus the NF is again linear, ${\dot w} = Aw$ in normalized coordinates.
\medskip

\underbar{{\it Case C3.}} If $\b_0 = 0$ but $\om \not= 0$, the eigenvalues of $A$ do again split in two groups, with $m$ eigenvalues $\la_j = i \om$, and $m$ eigenvalues $\la_j = - i \om$. Now the eigenvalues do not belong to a Poincar\'e domain, and moreover they give origin to an infinite number of resonances.

However, now ${\cal C} (T)$ is {\it not} abelian, so the system is in
general {\it not} already in NF. To be more specific, it is better to use the
possibility to assume $A = \b_0 I + \om J$ with $J=K_\a$ (e.g. $J = K_1$), see remark 12; equivalently, consider $A = \b_0 I + \om J$, where $J =
\sum_\a (\b_\a / \om ) K_\a $. In the subcase we are considering, $\b_0 = 0$.

Now, $A^+ = - A$, and the most general matrix in ${\cal C} (T)$ which
commutes with this $A$, as required by normal forms\footnote{Notice that therefore $G$-symmetric systems [$G=SU(2)$] in normal form have a symmetry $\^G = SU(2) \times SO(2)$, greater than generic $G$-symmetric systems.} theory, is a linear combination of $I$ and $J$. From this it
follows at once that the NF for (7) -- with $\b_0 = 0$ and $\om \not= 0$ --
is of the form
$$ {\dot w} \ = \ \om \, J w \ + \ \sum_{k=1}^\infty \, r^{2k} \, [ a_k I +
b_k J ] \, w \ . \eqno(8) $$ 

Here again the linear part ${\dot x} = Ax$ represents a hamiltonian
system, while the full system (7) is in general -- i.e. unless $\a_0 (r^2)
\equiv 0$ and $\a_\b (r^2) = c_\b P (r^2)$ for a single scalar function
$P$ -- non hamiltonian; the NF is also in general non-hamiltonian, unless
all the $a_k$ vanish.

Again if the $a_k$ are not all zero, we can ``further normalize'' this NF; the formal computations are exactly the same as in the $SO(2)$ case, and we arrive at the same result, i.e. the statement of the
above Lemma 2. Notice that in this case the initial system has first to be
put in standard NF, and that the convergence of the normalizing
transformation for this first step is not guaranteed a priori, so that the convergence of the renormalizing transformation would 
not in itself guarantee that the initial system is actually conjugated to the renormalized form in any neighbourhood of the origin.
\medskip

{\it Remark 13.} The interpretation in terms of ``potentials'' (or
``Hamiltonians'') $\H_i$ is immediate: the system (8) can be written in terms of four of them, ${\dot x} = \sum_i K_i \nabla \H_i (r^2)$; in case C1 we can always normalize so that $\H_0 = (\b_0 /2) |w|^2$ and $\H_\a \equiv 0$, in case C2 we can always obtain $\H_0 = (\b_0 /2) |w|^2$, $\H_\a = c_\a
\H_0$, while in case C3 we are guaranteed to get this situation only
formally. $\odot$

\section{Convergence issues.}

It should be stressed that the above discussion has been conducted at the formal level, i.e. the series defining the normalizing changes of coordinates are in general only formal series. 

The problem of convergence (in a suitably small, but nonzero, neighbourhood of the origin) of these series, i.e. the problem of the relation between formal conjugacy and analytic conjugacy, is in general a very hard one and a very limited number of general results is available to guarantee convergence (or divergence) for classes of equations and normal forms; see \cite{GSPT} for a brief summary of available results, and \cite{CWt} for the role of symmetries in these issues.

In the context of the present discussion, we recall three results: 
\medskip

$\bullet$ {\it Poincar\'e criterion} \cite{Arn,AIl}: if the convex hull (in the complex plane) of eigenvalues of the linear part $A$ does not include the origin, then the system is analytically conjugated to its linear part. 

$\bullet$ {\it Sternberg theorem} \cite{AIl,Be1}: if the linear part $A$ of the system is hyperbolic, then formally conjugacy to the normal form implies $C^\infty$ (but in general not analytic) conjugacy. The theorem was proved by Sternberg and Chen \cite{Che,Ste}, and recently extended to symmetric systems \cite{BeK,BK2}.

$\bullet$ {\it BMW-C theory}: if the linear part of the system satisfies a very general arithmetic condition known as ``condition $\om$'' (introduced by Bruno \cite{Bru} building on work by Siegel and Pliss \cite{AIl,Pli}) and the normal form satisfies ``condition A'', i.e. can be written as ${\dot x} = [1 + \a (x) ] Ax$, where $\a (x)$ is a polynomial scalar function (with $\a (0) = 0 $), then the system is analytically conjugated to the normal form. The first idea in this direction was by Markhashov; the theory was then corrected and developed by Bruno and Walcher \cite{Mar,BrW}, and later extended by Cicogna and Walcher \cite{Cic,Wal2}; see also \cite{CG3,CWt}.
\medskip

It should be noted that in all cases but B3 and C3, the Poincar\'e criterion applies and thus we can guarantee convergence of the normalizing transformation (in a sufficiently small neighbourhood of the origin) in general terms.

In cases B3 and C3 neither the Poincar\'e criterion nor the Sternberg theorem can apply. In general,  condition A is also not satisfied; however, in special cases  the normal form can satisfy condition A, and in this case convergence is guaranteed.
\medskip

{\it Remark 14.} In this note we are not considering systems symmetric under reducible representations, but I would like to stress that for reducible representations which are the sum of irreducible representations of the types A, B1, B2, C1 and C2, the Poincar\'e criterion is in general not satisfied (eigenvalues corresponding to different blocks are independent and can have real part of opposite signs), but the conditions for Sternberg theorem still hold: thus we are guaranteed of $C^\infty$ conjugacy of the system with its normal form. In many physical situations -- e.g. when the series (1) corresponds to a Taylor expansion -- this can be as interesting as analytic conjugacy. $\odot$
\medskip

{\it Remark 15.} In case C3 one should expect the normal form transformation to be generically nowhere convergent: indeed the original system has a $SU(2)$ symmetry, while the system in normal form has a $SU(2) \times U(1)$ symmetry (the additional $U(1)$ factor corresponding to commutation with the linear part); the two are hence qualitatively different and should not be expected to be conjugated by a diffeomorphism. $\odot$

\section{Vanishing linear part}

As stressed in the Introduction and in other occurrences, the standard normal forms theory, based on the homological operator $\L_0 := \{ f_0 , . \}$ associated to the linear part of the system, is not able to cope with systems having vanishing linear part. This limitation is not suffered  by Poincar\'e renormalized forms (PRF), where we use also higher order homological operators associated to higher order parts of the system, $\L_k := \{ f_k , . \}$. In this section we apply PRF  theory \cite{CG3,Ga2} to systems with $A=0$ in the different cases considered above.

We refer to \cite{CG3,Ga2,Ga4,GaB} for details on PRF procedure; here it will be enough to recall that for each terms $f_k$ we can eliminate the parts in ${\rm Ran} (\M_0) \cup  ... \cup {\rm Ran} (\M_{k-1})$, where $\M_s$ is the restriction of $\L_s$ to the kernel of the $\L_p$ with $p<s$, and $\M_0 = \L_0$.

It should be noted that the set of vector fields in normal form with respect to a given (semisimple) linear part is always a Lie algebra; when the algebraic structure of this is favourable, one can employ it to get a better reduction than with the generic PRF algorithm (still employing the $\L_k$ operators); in this case one speaks also of ``Lie renormalized forms'' (LRF) \cite{GaB,GLie}. 

Not surprisingly in view of the fundamental role of symmetry in our discussion, this is indeed the case for the situation we are analyzing.

\medskip
 
\subsection{Case A: $C(T) \simeq \R$}
 
Here the general form allowed by symmetry is given by (5); we rewrite this in the form of a power series expansion as
$ {\dot x}  =  \sum_{k=0}^\infty a_k \Psi_k $, where
$\Psi_k = r^{2k} x$. These satisfy $\{ \Psi_k , \Psi_m \} = 2 (m-k) \Psi_{k+m}$.
 
Let us now assume that $a_k = 0$ for $k < \mu$ and $a_\mu \not= 0$, with $\mu \ge 1$ (or we would have a linear part). Acting with $\L_\mu$ on generators $h_k$ for the (changes of coordinates given by) Lie-Poincar\'e transformations allowed by symmetry, which are necessarily of the form $h_k = \b \Psi_k$, will give
$$ \L_\mu (h_k) \equiv \{ a_\mu \Psi_\mu , \b \Psi_k \}  =  2 a_\mu \b (k-\mu) \Psi_{k+\mu} \ . \eqno(9)$$
We can therefore eliminate all the higher order terms except $\Psi_{2 \mu}$; the renormalized form results to be
$$ {\dot w} \ = \ (a_\mu r^{2 \mu} + \a r^{4 \mu} ) \, w \eqno(10) $$
where $\a$ is a real number, in general different from $a_{2 \mu}$.
\medskip

\subsection{Case B: $C(T) \simeq \C$}

Here the general form allowed by symmetry is given by (6); expanding it as a power series we get
$ {\dot x} = \sum_{k=0}^\infty (a_k \Psi_k + b_k \Phi_k )$, where $\Psi_k$ is as above and $\Phi_k = r^{2k} J x$.

We now have $\{ \Psi_k, \Psi_m \} = 2 (m-k) \Psi_{k+m}$, $\{ \Phi_k , \Phi_m \} = 0 $, and $\{ \Psi_k , \Phi_m \} = 2 m \Phi_{k+m}$. 

We assume that $a_k = 0 $ for $k < \mu$ and $a_\mu \not= 0$, and $b_k = 0 $ for $k < \nu$ and $b_\nu \not= 0$, with $\mu \ge 1$ and $\nu \ge 1$ (or we would have a nonzero linear part). Here we should distinguish the cases $\mu < \nu$, $\mu = \nu$ and $\mu > \nu$. The generators of the transformations should respect the symmetry, i.e. be of the form $h_k = \a_k \Psi_k + \b_k \Phi_k$.

For $\mu < \nu$, it follows from the commutation relations given above that we can eliminate all terms of the form $\Phi_{\mu +k}$, and all the terms $\Psi_{\mu+k}$ at the exception of $k = \mu$. Thus the renormalized form is in this case
$$ {\dot w} \ = \ \left[ r^{2 \mu} ( a_\mu  + \a r^{2 \mu} ) \right] \ w \eqno(11) $$
where $\a$ is a real number, in general different from $a_{2 \mu}$.

For $\nu < \mu$, one can easily eliminate all terms $\Phi_{\mu + k}$ 
and, again, all terms $\Psi_k$ except those for $k = \mu$ and $k = 2 \mu$; but not the terms $\Phi_m$ for $\nu \le m \le \mu$. 

Notice that this is obtained eliminating first the $\Psi_{\mu + k}$ terms by the choice of $\a_k$ in $h_k$, and then the $\Phi_{\mu + k}$ by the choice of $\b_k$ in $h_k$; that is, we are employing the Lie algebra structure of vector fields in normal form (the LRF algorithm \cite{GLie}).

Thus the renormalized form is in this case
$$ {\dot w} \ = \ \left[ r^{2 \mu} ( a_\mu  + \a r^{2 \mu} ) \, I \ + \ \sum_{k=\nu}^\mu \b_k r^{2 \nu} J \right] \ w \eqno(12) $$
where $\a$ and the $\b_k$'s are real numbers.
\medskip

\subsection{Case C: $C(T) \simeq {\bf H}$}

The general form allowed by symmetry is now, with the notation introduced in sect. 3.3, ${\dot x} = \sum_{p=0}^3 \a_p (r^2) K_{(p)} x$; expanding it in a power series we get 
$$ {\dot x} \ = \ \sum_{k=0}^\infty r^{2k} \left[ a_k I + \sum_{p=1}^3 \, b^{(p)}_k K_{(p)} \right] \ x \ . \eqno(13) $$

We introduce $\Psi_k$ as above, and $\Phi^{(p)}_k = r^{2k} K_{(p)} x$, so that we have 
$$ {\dot x} \ = \ \sum_k \left( a_k \Psi_k + \sum_p b_k^{(p)} \Psi^{(p)}_k \right) \ . \eqno(14)$$
Note that the $\Psi$ and $\Phi$ satisfy the commutation relations
$$ \begin{array}{l}
\{ \Psi_k , \Psi_m \} \ = \ 2 (m-k) \Psi_{k+m} \\ 
\{ \Psi_k , \Phi^{(p)}_m \} \ = \ 2m \Phi^{(p)}_{k+m} \\ 
\{ \Phi^{(p)}_k , \Phi^{(q)}_m \} \ = \ 2 \epsilon_{pqs} \Psi^{(s)}_{k+m} \end{array} $$ where $\epsilon_{pqs}$ is the completely antisymmetric (Levi-Civita) tensor on three indices. 

We assume that $a_k = 0 $ for $k < \mu$, $a_\mu \not= 0$, and that $b_k^{(p)} = 0 $ for $k < \nu_{(p)}$, $b_{\nu_p} \not= 0$, as usual with $\mu , \nu_{(p)}$ positive (or we would have a nonzero linear part).

One should distinguish several cases depending on the relations between $\mu$ and $\nu_{(p)}$. The normalizing transformations generators should respect the symmetry, i.e. be of the form
$ h_k = \a_k \Psi_k + \sum_p \b_k^{(p)} \Phi^{(p)}_k $. We will again use the Lie algebraic properties of vector fields in normal form. 

For $\mu < \nu_{(p)}$ ($p=1,2,3$), we can eliminate all terms $\Phi_k^{(p)}$, and all terms $\Psi_{\mu+k}$ with $k \not= \mu$.
If some of the $\nu_{(p)}$ are equal to $\mu$, the corresponding term $\Phi_\mu^{(p)}$ cannot be eliminated. Thus the renormalized form is in this case
$$ {\dot w} \ = \ \left[ r^{2 \mu} (a_\mu + \a r^{2 \mu} ) I \, + \, \sum_{p=1}^3 \b_\mu^{(p)} K_{(p)} \right] \ w \ \ \ \ \ (\mu \le \nu_{(p)}) \eqno(15)$$
where $\a$ and the $\b_{(p)}$ are real numbers, and $\b_{(p)} = 0 $ if $\mu < \nu_{(p)}$.

If $\nu_{(s)} < \mu$ and $\nu_{(s)} < \nu{(p)} $ for $p \not=s$ (we write simply $\nu$ for $\nu_{(s)}$), then the same considerations as in the previous case B applies. 
Thus, the renormalized form is in this case
$$ {\dot w} \ = \ \left[ \left( a_\mu r^{2 \mu} + \a r^{4 \mu} \right) \, I \ + \ \sum_{k=\nu}^\mu \ \sum_{p=1}^3 \ \b_k^{(p)} r^{2 \nu_{(p)}} K_{(p)} \right] \ w \ . \eqno(16)$$
Here it is again essential to use the LRF algorithm \cite{GLie}.

\section{Conclusions.}

We have considered Poincar\'e-Dulac normal forms for dynamical systems
(vector fields, formal power series) in $\R^n$ in the presence of a symmetry group $G$ with Lie algebra $\G$ when $G$ is a simple Lie groups (actually any group with transitive action on $S^{n-1} \ss \R^n$) and thus $\G$ a simple Lie algebra of the $A - G$ types, acting through its fundamental (defining) real representation. We have used the fact that in this case the equivariant dynamical systems (vector fields, formal power series) are of the form (4).

We have completely described the resulting normal form, which moreover are guaranteed to be obtained by a transformation which is convergent in some neighbourhood of the origin in all subcases but two (B3 and C3, in the latter we argued convergence should not be expected); in some cases the symmetry guarantees the system is necessarily in normal form without the need of any transformation.

In the cases B3 and C3 the normal form contains an infinite number of terms, but it is possible to ``further normalize'' them (pass to ``Poincar\'e renormalized forms''  \cite{Ga2}, or better to ``Lie renormalized forms'' \cite{GaB,GLie}) obtaining a much simpler expression, except when we are in case B3 and the system is hamiltonian; this case is covered by other approaches, and the same result applies \cite{FM,SM}.

\end{document}